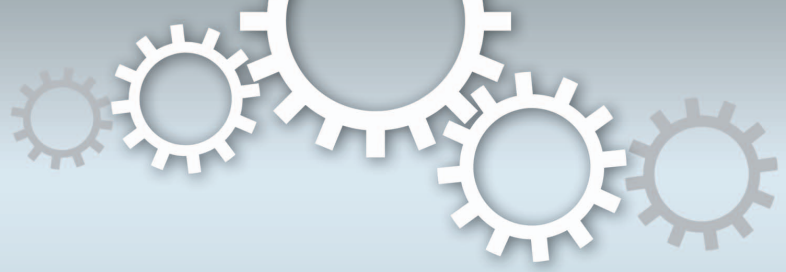

**OPEN**

# A Unifying Framework for Measuring Weighted Rich Clubs

Jeff Alstott[1,2], Pietro Panzarasa[3], Mikail Rubinov[1,4], Edward T. Bullmore[1] & Petra E. Vértes[1]

[1]Department of Psychiatry, Behavioural and Clinical Neuroscience Institute, University of Cambridge, Cambridge CB2 0SZ UK, [2]Section on Critical Brain Dynamics, National Institute of Mental Health, Bethesda, 20892 Maryland, USA, [3]School of Business and Management, Queen Mary University of London, London, E1 4NS UK, [4]Churchill College, University of Cambridge, Cambridge CB2 0SZ UK.

Network analysis can help uncover meaningful regularities in the organization of complex systems. Among these, rich clubs are a functionally important property of a variety of social, technological and biological networks. Rich clubs emerge when nodes that are somehow prominent or 'rich' (e.g., highly connected) interact preferentially with one another. The identification of rich clubs is non-trivial, especially in weighted networks, and to this end multiple distinct metrics have been proposed. Here we describe a unifying framework for detecting rich clubs which intuitively generalizes various metrics into a single integrated method. This generalization rests upon the explicit incorporation of randomized control networks into the measurement process. We apply this framework to real-life examples, and show that, depending on the selection of randomized controls, different kinds of rich-club structures can be detected, such as topological and weighted rich clubs.

The mesoscopic organization of networks is often described in terms of communities of densely connected elements, such as the trading blocs within the world economy[1]. These communities are often regarded as the functional subunits of the network[2,3], a perspective that is premised on a somewhat egalitarian view of a network's organization that does not differentiate between individual nodes in terms of their functional importance. Yet some nodes are more influential than others[4,5]. For instance, to fully understand the coordination of global trade, it would be necessary to recognize the crucial role of global cities, prototypically New York[6]. Such 'important' cities may be in separate communities, but together they exercise an enormous influence over the transactions occurring in the global economic system. The study of interactions between the important or prominent elements in the network is therefore useful for understanding its mesoscopic organization. Here we focus on one kind of interaction between the prominent network elements, in which they organize into subgroups called rich clubs. We catalogue and assess various proposed metrics for measuring rich clubs in weighted networks. We create a framework that generalizes and unifies these disparate metrics and, in many cases, simplifies their calculation.

Rich clubs are subgroups of important or influential (rich) nodes that preferentially interact with one another[7,8]. This network feature has been detected in diverse complex systems, including transportation networks, scientific collaboration networks, and the human brain[8,9]. Rich clubs can serve as a network's backbone for optimizing the routing between peripheral nodes[10], and can substantially affect a number of the network's properties, including clustering and assortativity[11]. Accordingly, targeted attacks to connections within the rich club can damage a network's connectivity more than attacks to the links of highly connected nodes that are not rich-club members[9].

Rich clubs may be topological or weighted. In topological rich clubs, the rich nodes preferentially create connections with each other[7,8]. In weighted rich clubs, the rich nodes preferentially allocate weight to the connections between each other[12]. The weight of a connection may represent its intensity, capacity, duration, intimacy, or exchange of services[13].

The topological rich-club coefficient $\phi$ is the ratio between the number of existing connections between the rich nodes, $E$, and the number of possible connections between them[7,8]. For a given set of $N$ rich nodes, the coefficient is formalized as:

$$\phi = \frac{2E}{N(N-1)}, \qquad (1)$$

where the number of possible undirected connections is $N(N - 1)/2$. Node 'richness' can be defined in terms of any property (the richness parameter $r$), insofar as all nodes can be ranked according to that property. Commonly, however, $r$ is taken to be node degree (i.e., the number of links incident upon a node). The rich nodes are those whose richness exceeds a threshold $\bar{r}$. The rich-club coefficient $\phi$ can thus be calculated in relation to any value of $\bar{r}$, and is commonly represented as a curve showing the value of $\phi$ within a range of thresholds $\bar{r}$ (as is done in examples below). The value $\phi$ is then normalized by comparison to the same value observed in randomized controls, $\phi_{rand}$:

$$\phi_{norm} = \frac{\phi}{\phi_{rand}}. \quad (2)$$

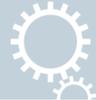

These controls have the same richness sequence as the real network (i.e., the set $r_1, r_2, r_3, \ldots$ of richness values for all the nodes)[8]. The normalization identifies whether the number of connections between the rich nodes exceeds what is expected simply by chance, and thus forms a club of preferentially allocated connections.

Despite the straightforward definition of the topological rich-club coefficient, its generalization to weighted rich clubs is not trivial. Numerous metrics for measuring weighted rich clubs have been proposed and implemented, each with distinct implicit assumptions. The diversity of these metrics and of their assumptions has received little explicit attention, even though each metric ultimately captures a different network property. The lack of a rigorous comparative assessment of metrics has, in turn, blurred the relation and consistency between different assessments of weighted rich clubs across a variety of empirical studies.

Here we describe the relationship between the individual metrics previously proposed and several new metrics we introduce. Importantly, we then generalize all these metrics by showing how they can be reduced to a common unifying framework. The generalization relies critically on integrating randomized controls into the measurement process. We then employ this framework to show how, by selecting different randomized controls, different kinds of rich-club structures can be captured. As an example, we illustrate how topological and weighted rich-club properties can be distinct and can be measured separately using different randomized controls. We discuss how methods that do not explicitly integrate randomized controls can inadvertently conflate these two rich-club structures.

## Results

**Integration of rich-club measurements.** A number of metrics that extend Eq. 1 to weighted networks have been proposed[8,12,14,15]. All these metrics have the form:

$$\phi = \frac{C}{F}, \quad (3)$$

where $C$ is the weighted connectedness of the club (Figure 1), defined as the sum of the weights of the links between the rich nodes, and $F$ is the maximal possible weighted connectedness of the club. The value $F$ depends on domain-specific assumptions about how weights and links could be added or redistributed across the network.

Figure 2 contains previously proposed as well as several novel measures of $F$, organized along two dimensions: i) how many links could contribute to $F$ (rows), and ii) where the weights associated to these links are drawn from (columns). For example, in the first row, given the four rich nodes in the network, the maximum number of links that could exist between them is set at $P = 6$. But what is the maximum weight these links could carry? In the first column, we assume that any link could have a maximal weight $W_{max}$. This simple extension of Eq. 1 is relevant for instance in correlation-based financial networks, where every pair of nodes (stocks) could in principle have highly correlated activity (strong connection), and where correlation values are capped at $W_{max} = 1$[16]. In the second column, we assume that weights are tied to the links, so that the maximal connectivity would be achieved by placing the six strongest links existing in the whole network inside the rich club. The third column is sim-ilar, except that we assume that only links already attached to one of the rich nodes can be rewired to serve intra-club connectivity.

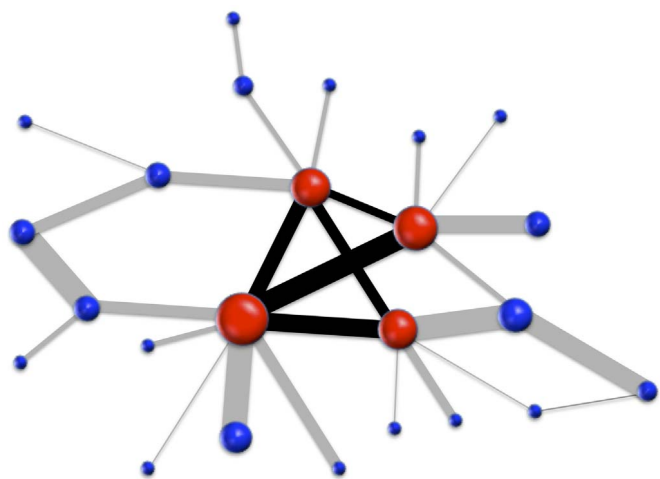

**Figure 1 | Weighted rich clubs are measured in terms of the weighted connectedness between rich nodes.** The rich club is the set of red (color online) nodes with a richness value above a certain threshold $\bar{r}$. In this example, the richness parameter is degree and $\bar{r} = 3$. Size of nodes is proportional to their richness. The rich club is thus the subgraph formed by nodes with degree larger than 3. The weighted connectedness, $C$, of the rich club is the sum of the weights of the links between the nodes in the subgraph (black lines). The rich-club coefficient $\phi$ is calculated by dividing the existing weighted connectedness $C$ by the maximal possible weighted connectedness, $F$. See Eq. 3 and Figure 2.

The second row in Figure 2 is similar to the first row, but assumes that the number of connections that exist among the rich nodes is fixed (at $E = 5$ in the example shown) and no new connections could be formed[12,17].

In contrast to the first two rows, the third row assumes that weights can be redistributed across the links of the network. For example, panel (i) implies that the total strength of the nodes (i.e., the sum of the weights of the nodes' links) is fixed, but rich nodes could redistribute their weights across their links. This would be appropriate for a social network in which individuals (nodes) could choose to redistribute the time spent with various friends (i.e., weight of links)[12]. Members of the rich club could therefore in principle choose to direct all their time (sum of all weights) toward interactions with other club members.

The use of a domain-specific denominator $F$ has been a common practice; yet, it is unnecessary, it may involve practical difficulties in its calculation, and it may induce biases in the detection of rich clubs by obscuring the influence of randomized controls. Indeed, the evaluation of the weighted connectedness of a rich club requires normalization through comparison to randomized controls with the same richness sequence, just as with unweighted rich clubs[12]. This normalization is required because even networks in which links are randomly established can exhibit a non-zero value of $\phi$. The detection of rich clubs in real networks must therefore take into account and discount how much weighted connectedness would be expected simply by chance[8,12,17]. Let us consider the unweighted rich-club metric given by Eq. 1. Because the richness sequence is preserved, the number of rich nodes $N$ at a given threshold $\bar{r}$ will remain constant. Thus:

$$\phi_{norm} = \frac{\phi}{\phi_{rand}} = \frac{2E}{N(N-1)} \bigg/ \frac{2E_{rand}}{N(N-1)} = \frac{E}{E_{rand}}. \quad (4)$$

Similarly, for weighted rich clubs, if $F$ is preserved in the randomized control, then $\phi_{norm}$ simplifies to:

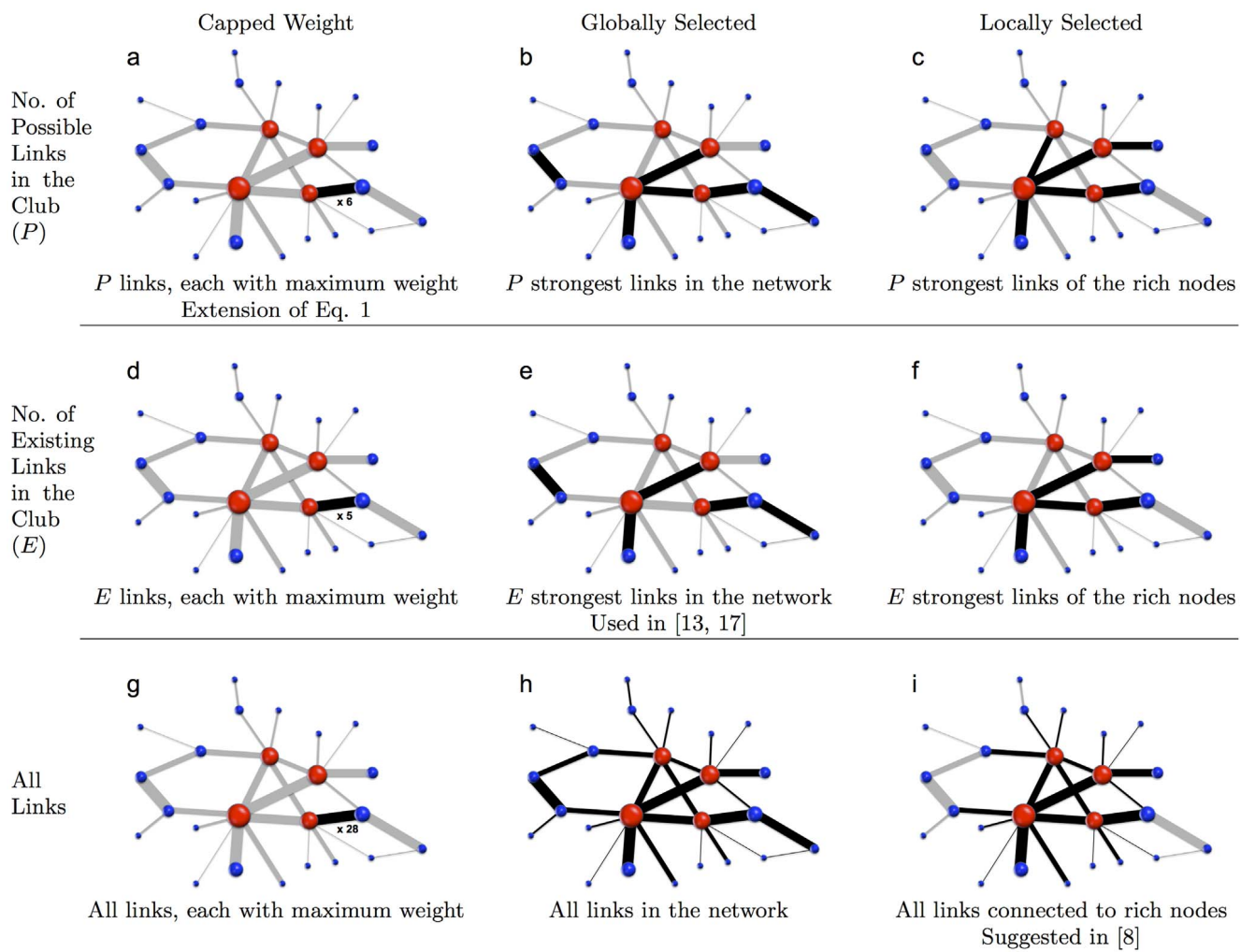

**Figure 2 | Nine ways to measure weighted rich clubs, which all simplify to Eq. 5.** The rich-club coefficient $\phi$ is calculated by comparing the existing weighted connectedness of the rich club to the maximal possible weighted connectedness, $F$. In each panel, the size of nodes is proportional to their richness, and the width of links to their weight. Red nodes are the members of the rich clubs. Each panel describes an alternate way to define the maximal weighted connectedness, and shows the set of links (black lines) whose collective weight is $F$. Underlying each metric are different assumptions about how links and weights could, in principle, be alternatively arranged in the network to yield the maximal possible weighted connectedness. Preserving these assumptions in the creation of randomized controls will ensure that the normalized rich-club coefficient $\phi_{norm}$ simplifies to Eq. 5. The nine measures in this figure are organized along two dimensions: (i) how many links could contribute to F (rows), and (ii) where the weights associated to these links are drawn from (columns). *Left Column (Capped Weight)* Assumes links could have any weight up to a specific maximal weight. *Middle Column (Globally Selected)* Assumes weights are attached to the links, so that the maximal weighted connectivity would be achieved by taking the strongest links from anywhere in the network and placing them inside the rich club. *Right Column (Locally Selected)* Assumes only links connected to rich nodes can be locally rewired to serve intra-club weighted connectivity. *First Row (P Links)* Assumes additional links can be added within the club, up to the topological limit $P$ ($P = 6$ in this example). *Second Row (E Links)* Assumes the number of links in the club is fixed at the existing number, $E$ ($E = 5$ in this example). *Third Row (All Links)* Assumes weights can be redistributed among the links of the network.

$$\phi_{norm} = \frac{\phi}{\phi_{rand}} = \frac{C}{F} \bigg/ \frac{C_{rand}}{F} = \frac{C}{C_{rand}}. \qquad (5)$$

The resulting simplified form of the rich-club metric in Eq. 5 has been used previously[18], though without the motivating arguments presented here. Here we argue that this simplified metric is in fact necessary, since other previously proposed metrics all simplify into this form when appropriate randomized controls are used. Indeed, appropriate randomized controls should impose the same constraints on the rewiring of links and weights, and thus have the same maximal weighted connectedness within the club $F$ as the real network. It then follows that any $\phi_{norm}$ will always yield Eq. 5, because the two values of $F$, one at the denominator of $\phi$ and the other at the denominator of $\phi_{rand}$, cancel each other out.

**Empirical applications.** The framework of Eq. 5, with the explicit inclusion of $C_{rand}$ into the rich-club metric, helps us focus our attention on the importance of selecting the appropriate randomized controls to detect rich clubs. Randomized controls ensure that the measured rich clubs are not a trivial consequence of connectivity constraints or other features of the network. In the case of topological rich clubs, such constraints could include the degree sequence (i.e., the set $k_1, k_2, k_3, \ldots$ of degrees for all the nodes) as well as more complex properties such as betweenness centrality. In the case of weighted rich clubs, these potentially confounding features of the network could include topological rich clubs originating only from link placement. The traditional use of randomized controls that only preserve the degree sequence of the real network can conflate the existence of a purely topological rich-club structure (due to link placement) with the existence of a weighted rich-club structure (due to weight allocation) based on a given network topology.

Here we use two empirical networks to illustrate the distinction between topological and weighted rich-club structures and cast light on the crucial role of randomized controls in the detection of one structure versus the other. We examined two networks: (i) the human brain, in which nodes are the cortex regions, links are the white matter tracts between regions, and weights are the number of fibers per tract[10] (Figure 3a, left); and (ii) the global airline traffic network, in which nodes are the airports, links are the routes between airports, and weights are the number of flights per route[12,17] (Figure 3a, right). These networks are concerned with different empirical domains and have different spatial scales; yet they both have been found to be organized into topological and weighted rich clubs, with widespread practical implications for their effectiveness and efficiency.

In neural systems, on the one hand, a distinctive functional significance has been attributed to rich clubs[24]. For instance, schizophrenia has been associated with altered rich-club structures[25]. Analysis of rich clubs in the brain has shown that they can facilitate the connection of peripheral nodes of the network, over and above the effect of high-degree hubs that are not members of a club[10]. Transportation systems, on the other hand, have been found to exhibit skewed distributions of travel fluxes per connection, as well as a pronounced heterogeneity in the traffic passing through various locations[26,27]. In particular, previous studies have suggested that traffic in airline networks tends to be channeled on the routes connecting the busiest airports[8,12,17]. This is responsible for the emergence of rich clubs in which large backbones of travel fluxes are controlled by the airports with many intercontinental connections to other hubs. The organization of airline transportation systems into rich-club structures, in turn, has a number of policy implications for the management and control of air traffic, ranging from the optimization of the operating costs for carriers and the average traveling time for passengers, to the resilience of the systems against unexpected failures and bottlenecks[28].

Both networks clearly exhibit topological rich clubs. As shown by Figure 3b (left), in the brain network, highly connected cortex regions tend to be linked with one another to a larger extent than would be expected by chance. However, when topology is controlled for, the weight allocation in the brain network shows a pattern that is the opposite of rich-club behavior, as indicated by Figure 3c (left). In this case, the links between high-degree nodes have less weight than would be expected by chance, in spite of the tendency of such nodes to connect with one another.

Similarly, in the airline network, highly trafficked airports, with routes to many destinations, tend to establish connections with one another to a larger extent than randomly expected (Figure 3b, right). However, unlike the brain network, the airline network is organized not only into topological, but also weighted rich clubs (Figure 3c, right). Over and above the propensity to direct routes toward one another, highly trafficked airports also tend to allocate more flights among one another than randomly expected. This finding is in qualitative agreement with previous work that has documented non-trivial correlations between degree and weight of links in transportation networks[26,27]. Because airports with many routes are also characterized by large travel fluxes per route, hub airports can secure and share control over such fluxes by preferentially directing their routes toward one another. One speculative explanation for why the brain and airline networks show opposite weighted rich-club behavior could be that the two systems have different goals: airline networks integrate regions, while brain networks must achieve both integration and segregation of processing regions. Thus in the airline network both link placement and link weight serve to increase connectivity between cities; in the brain link placement also increases connectivity, but lower link weight keeps processing regions functionally distinct.

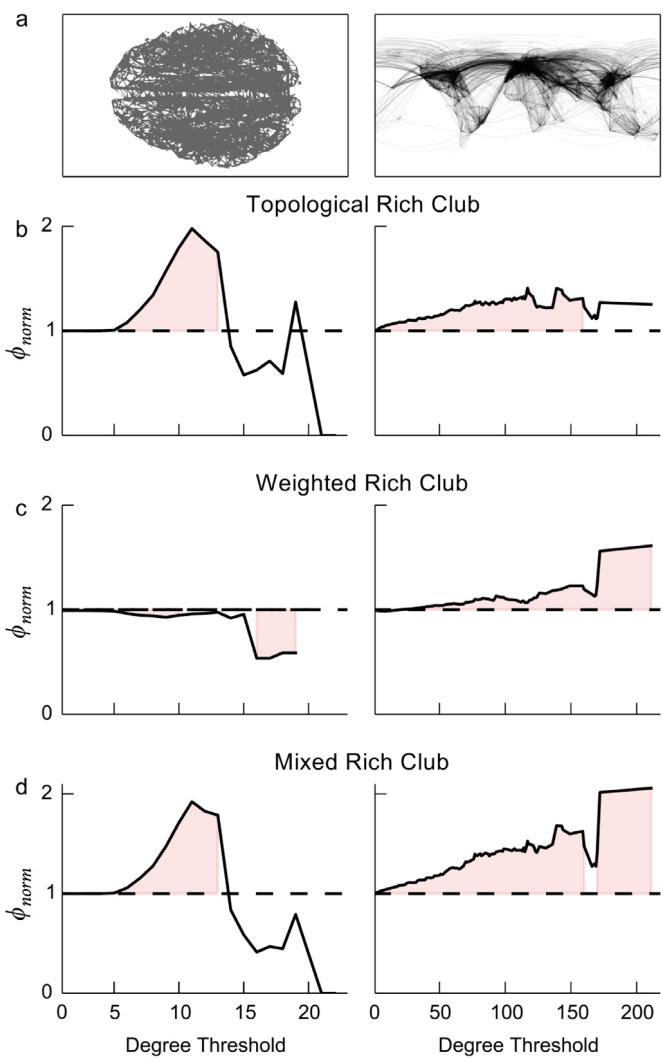

Figure 3 | Topological and weighted rich-club structures intermingle in networks, but can be distinguished by using different randomized controls. (a) Visualization of the two weighted networks examined: the white matter network of the human brain (left, data from[10], link weight: number of fiber tracts) and the global commercial airline network (right, data from[21], link weight: flights per day). (b–d) The rich-club behavior as measured by applying different metrics to the white matter network of the human brain (left) and the global airline network (right). $\phi_{norm}$ is the normalized rich-club coefficient, based on 1, 000 randomized controls. The shaded areas highlight those values for which $\phi_{norm}$ is significantly different from 1 ($p < .05$). Rich-club coefficients were calculated for every unique value of degree in the network. (b) The topological (unweighted) rich-club behavior, based on randomized controls with the same degree sequence as the real network. Both networks show topological rich clubs, in qualitative agreement with what was found in[8,10,12]. (c) The weighted rich-club behavior, based on randomized controls characterized by the same topology as the real network, but with uncorrelated weights. (d) The mixed rich-club behavior, based on controls that randomize the topology and also decorrelate the weights. Similar controls were used to measure weighted rich clubs in the networks analyzed in[10].

Further research could examine how these distinct topological and weighted rich-club structures yield different system behaviors.

For both the brain and the airline networks, measuring the distinctive contributions of topology and weights to rich-club structures requires the use of the appropriate null models. If randomized controls that alter both topology and weights are used, the two contributions are inevitably mixed and combined, and thus the two distinct types of rich-club structures cannot be properly disentangled. As shown by Figure 3d (left), the brain network exhibits a mixed rich-club structure. In this case, the negative contribution of weights is more than compensated for by the positive contribution of topology to the rich clubs. The resulting net effect is a rich-club ordering that is not unambiguously attributable to either topology or weights, but only to some combination of the two sources. Similar mixed evidence is shown with the airline network (Figure 3d, right). In this case, the network exhibits a more pronounced tendency toward a mixed rich-club structure than toward topological or weighted rich clubs. While in the brain network topology and weights have distinct effects of different sign, in the airline network both effects have a positive sign and combine into synergies yielding a compound and mixed rich-club ordering.

Mixed rich clubs may be measured intentionally (as in Figure 3d). However, the previous framework of Eq. 3 ($\phi = C/F$) made unintentional (and inaccurate) measurements of these mixed patterns more likely to occur as a result of mismatches between $F$ and the randomized controls. A number of previous studies have reported weighted rich clubs based on $\phi = C/F$, where the quantity $F$ was the sum of the $E$ strongest links in the network, as shown in Figure 2[9,10,12,17]. This quantity assumes that the number of links in the club is fixed, at $E$, and as such is suitable to measuring solely the behavior of weights and their contribution to rich-club structures, above and beyond the contribution of topology. However, to calculate $\phi_{norm}$, $\phi$ was then compared to randomized controls that shuffled link placement, and thus changed the number of links within the club[9,10]. This introduced a mismatch between the assumption underlying the quantity $F$ and the one underlying the randomized controls. For $F$, the assumption is that the topology is fixed, so that the maximum number of links that could exist between members of the rich club is fixed at the existing number of links within the club, $E$. In contrast, for the randomized controls the assumption is that the number of links existing within the club is allowed to vary. By reintroducing the effect of link placement on rich clubs through the randomized control, the value ultimately calculated inadvertently conflated topological and weighted rich-club structures. Such potential mismatches between $F$ and randomized controls are circumvented by our framework of Eq. 5 ($\phi_{norm} = C/C_{rand}$), which eliminates the need to calculate $F$ and places emphasis on the selection of randomized controls.

## Discussion

Here we have generalized previously reported metrics for measuring weighted rich clubs, and created a unifying framework based on the direct inclusion of randomized controls. Just as the various metrics in Figure 2 were originally proposed to address conceptually different questions, the present framework enables different properties of a network to be evaluated through different randomized controls. For example, we have shown that purely topological rich clubs can be measured using controls that shuffle topology, while weighted rich clubs can be measured using controls that preserve topology but decorrelate weights.

More complex measurements and controls can be envisaged, and a number of procedures have been proposed to create different types of controls. For instance, the strength sequence (i.e., the set $s_1, s_2, s_3, \ldots$ of the strengths of all the nodes) of a given network can be preserved while uncorrelated values can be assigned to the weights of the links[19]. It is also possible to preserve the weight sequence (i.e., the set $w_1, w_2, w_3, \ldots$ of the weights attached to all the links), for networks with many links[20]. Moreover, for directed networks, the out-strength or in-strength sequence could be preserved by randomizing the weights of each individual node's outgoing or incoming links, respectively[12]. This procedure has the added effect of preserving not only the weight sequence for the entire network, but also the out-weight or in-weight sequence of each individual node (i.e., the set of weights of all links departing from, or arriving at, each node).

Even more sophisticated properties could be preserved in randomized controls, such as network structural features (e.g., betweenness centrality) or non-network features (e.g., the demographic attributes of individuals in a social network). These controls can then be used to isolate any contribution those features make to the connectedness of a rich club, and thus to assess whether any rich-club structure remains. Ultimately, the choice of the appropriate randomized control requires domain-specific information as to how links in a given network could in principle be added, severed or reshuffled, and the weights reallocated to existing or newly established links.

In summary, while a number of metrics have been suggested for uncovering topological and weighted rich-club structures[8,12,14,15], the framework described here simplifies these metrics to a single form by explicitly incorporating randomized controls into the measurement process. In doing so, this framework lays bare how to disentangle and measure the distinct contributions of topology and weight assignment to the generation of rich-club structures.

## Methods

Rich-club effects were measured in two empirical networks: the human brain and the global commercial airline network.

The human brain network, from[10], is the white matter connectivity network of the cortex, as measured with diffusion tensor imaging data from 40 healthy subjects. The cortex was divided into 1,170 equally sized regions, which served as the nodes of the network. The links between the nodes are the white matter tracts between cortex regions, and link weights are the number of fibers found for each connection.

The global commercial airline network, from[21], was created from the openflights.org airline route database as of August, 2011. Network nodes are airports, links are flight routes between airports, and link weights are the number of flights per route.

For each network, the rich-club coefficient $\phi_{norm}$ was calculated using Eq. 5, and $C_{rand}$ was calculated from the average of 1,000 randomized controls. To measure topological rich clubs, link weights were ignored (i.e., they were set to one) and randomized controls with the same degree sequence were generated using the Viger-Latapy method[22], as implemented in the igraph toolbox[23]. To measure the weighted rich clubs, the link weights were included, and randomized controls were generated with the same topology but new, uncorrelated link weights. Uncorrelated link weights that closely preserved the strength sequence were created using the methods by Serrano et al.[19]. Lastly, to measure mixed rich clubs, both methods were combined, using randomized controls with both shuffled topology and uncorrelated weights.

1. Barigozzi, M., Fagiolo, G. & Mangioni, G. Identifying the community structure of the international-trade multi network. *Physica A* **390**, 2051–2066 (2011).
2. Girvan, M. & Newman, M. E. J. Community structure in social and biological networks. *Proc. Natl. Acad. Sci. U.S.A.* **99**, 7821–7826 (2002).
3. Fortunato, S. Community detection in graphs. *Phys. Rep.* **486**, 75–174 (2010).
4. Albert, R., Jeong, H. & Barabási, A. Error and attack tolerance of complex networks. *Nature* **406**, 378–382 (2000).
5. Guimerà, R. & Nunes Amaral, L. A. Functional cartography of complex metabolic networks. *Nature* **433**, 895–900 (2005).
6. Sassen, S. *The Global City: New York, London, Tokyo* (Princeton University Press, 2001).
7. Zhou, S. & Mondragón, R. J. The rich-club phenomenon in the Internet topology. *IEEE Commun. Lett.* **8**, 180–182 (2004).
8. Colizza, V., Flammini, A., Serrano, M. A. & Vespignani, A. Detecting rich-club ordering in complex networks. *Nature Phys.* **2**, 110–115 (2006).
9. van den Heuvel, M. P. & Sporns, O. Rich-club organization of the human connectome. *J. Neurosci.* **31**, 15775–15786 (2011).
10. van den Heuvel, M. P., Kahn, R. S., Goñi, J. & Sporns, O. High-cost, high-capacity backbone for global brain communication. *Proc. Natl. Acad. Sci. U.S.A.* **109**, 11372–11377 (2012).
11. Xu, K.-K., Zhang, J. & Small, M. Rich-club connectivity dominates assortativity and transitivity of complex networks. *Phys. Rev. E* **82** (2010).
12. Opsahl, T., Colizza, V., Panzarasa, P. & Ramasco, J. J. Prominence and control: The weighted rich-club effect. *Phys. Rev. Lett.* **101**, 168702 (2008).
13. Granovetter, M. S. The strength of weak ties. *Am. J. Soc.* **78**, 1360–1380 (1973).
14. Zlatic, V. *et al.* On the rich-club effect in dense and weighted networks. *Eur. Phys. J. B* **67**, 271–275 (2009).
15. Valverde, S. & Solé, R. Self-organization versus hierarchy in open-source social networks. *Phys. Rev. E* **76**, 046118 (2007).
16. Vértes, P. *et al.* Topological isomorphisms of human brain and financial market networks. *Front. Syst. Neurosci.* **5** (2011).
17. Ramasco, J., Colizza, V. & Panzarasa, P. Using the weighted rich-club coefficient to explore traffic organization in mobility networks. In *Lecture Notes of the*

Institute for Computer Sciences, Social Informatics and Telecommunications Engineering, vol. 4, 680–692 (Springer, Berlin Heidelberg, 2009).
18. Serrano, M. Rich-club vs rich-multipolarization phenomena in weighted networks. *Phys. Rev. E* **78** (2008).
19. Serrano, M. A., Boguna, M. & Pastor-Satorras, R. Correlations in weighted networks. *Phys. Rev. E* **74** (2006).
20. Rubinov, M. & Sporns, O. Weight-conserving characterization of complex functional brain networks. *NeuroImage* **56**, 2068–2079 (2011).
21. Opsahl, T., Agneessens, F. & Skvoretz, J. Node centrality in weighted networks: Generalizing degree and shortest paths. *Social Networks* **32**, 245–251 (2010).
22. Viger, F. & Latapy, M. Random generation of large connected simple graphs with prescribed degree distribution. *Computing and Combinatorics* **3595**, 440–449 (2005).
23. Csardi, G. & Nepusz, T. The igraph software package for complex network research. *InterJournal, Complex Systems* 1695 (2006).
24. Towlson, E. K., Vértes, P. E., Ahnert, S. E., Schafer, W. R. & Bullmore, E. T. The rich club of the C. elegans neuronal connectome. *J. Neurosci.* **33**, 6380–6387 (2013).
25. van den Heuvel, M. P. *et al*. Abnormal Rich Club Organization and Functional Brain Dynamics in Schizophrenia. *JAMA Psychiatry* **70**, 783–792 (2013).
26. Barrat, A., Barthélemy, M., Pastor-Satorras, R. & Vespignani, A. The architecture of complex weighted networks. *Proc. Natl. Acad. Sci. U.S.A.* **101**, 3747–3752 (2004).
27. Guimerá, R., Mossa, S., Turtschi, A. & Amaral, L. A. N. The worldwide air transportation network: Anomalous centrality, community structure, and cities' global roles. *Proc. Natl. Acad. Sci. U.S.A.* **102**, 7794–7799 (2005).
28. Barthélemy, M. Spatial Networks. *Phys. Rep.* **499**, 1–101 (2011).

### Acknowledgments
We would like to extend our gratitude to Vittoria Colizza for her comments and suggestions on the development of this paper. We would like to thank Tore Opsahl and Martijn van den Heuvel for making their data easily accessible. J.A. is supported by the NIH-Oxford-Cambridge Scholarship Program. P.P. is employed by Queen Mary University of London. M.R. is supported by the NARSAD Young Investigator and Isaac Newton Trust grants. E.T.B. is employed half-time by the University of Cambridge, UK, and half-time by GlaxoSmithKline (GSK). P.E.V. is supported by the Medical Research Council (grant number MR/K020706/1).

### Author contributions
J.A., P.V. and P.P. analyzed existing metrics. J.A., P.V. and M.R. created new metrics. J.A. and P.V. analyzed data. J.A., P.P., M.R., E.B. and P.V. wrote and reviewed the manuscript.

### Additional information
**Competing financial interests:** The authors declare no competing financial interests.

**How to cite this article:** Alstott, J., Panzarasa, P., Rubinov, M., Bullmore, E.T. & Vértes, P.E. A Unifying Framework for Measuring Weighted Rich Clubs. *Sci. Rep.* **4**, 7258; DOI:10.1038/srep07258 (2014).